# Kramers barrier crossing as a cooling machine

Philip R. Schiff and Abraham Nitzan

School of Chemistry, The Sackler Faculty of Science, Tel Aviv University, Tel Aviv, 69978, Israel

## Abstract

The achievement of local cooling is a prominent goal in the design of functional transport nanojunctions. One generic mechanism for local cooling is driving a system through a local uphill potential step. In this paper we examine the manifestation of this mechanism in the context of the Kramers barrier crossing problem. For a particle crossing a barrier, the local effective temperature and the local energy exchange with the thermal environment are calculated, and the coefficient of performance of the ensuing cooling process is evaluated.

# 1. Introduction

Studies of electronic transport in nanojunctions often involve issues of device stability and integrity, implying the need to consider heating and heat conduction in such systems.[1] In addition to technological implications, these considerations raise fundamental questions concerning heat generation and dissipation in driven nanosystems.[2-3] While energy is globally released in such driven processes, local cooling may be achieved in parts of the system, as was recently discussed[4-7] and possibly observed.[8] The underlying mechanisms for such cooling phenomena may be broadly divided into three classes. In one, energy dependent carrier fluxes distort the thermal distribution in the emitting electrode, potentially reducing its temperature if transport is biased towards higher energy carriers.[9-10] Thermoelectric cooling[11] belongs to this class as do some normal metal-insulator-superconductor junctions where cooling is effected by the favorable energy selection caused by the anisotropic density of states of the junction.[12-14] Another mechanism invokes charging induced capacitive forces to damp energy out of a bridge oscillating between two (source and drain) electrodes and controlled by an electrostatic potential imposed by a third (gate) electrode.[15-18] In the third mechanism, the system is driven through a local uphill potential and transport in this locality is facilitated by extracting heat from environmental modes. Laser cooling is a prominent example for this class of processes,[19-21] where the system is driven by light absorption and the uphill step is tailored by tuning the light frequency a little below resonance absorption. Analogous processes in conduction junctions use light-assisted transport in a similar manner.[22-23] However, because in such systems driving is provided by the voltage bias, electromagnetic modulation constitutes just a control tool and cooling may in principle be achieved without it if the intrinsic level structure of the bridging system provides the needed uphill step.[4]

This paper deals with the last mechanism, where local cooling is achieved by pushing a system through a local uphill step. Standard manifestations of this scheme, e.g. laser cooling by sub-resonance excitation, involve systems with discrete spectra coupled to the driving field and to their thermal environment. Here, we analyze this phenomenon withing the simplest classical model of this type, based on the Kramers barrier-crossing process. In Sec. 2, we recall the Kramers model for the barrier controlled dynamics of a particle transversing a barrier. We focus on the neighborhood of the barrier top since this is where non-equilibrium effects, which may lead to local cooling, dominate. The local effective temperature and heat exchange with the environment are evaluated in Sec. 3. In Sec. 4, we calculate the efficiency of this cooling process as the ratio between the rate of heat absorption from the environment in the cooling part of the process and the rate of energy input needed to keep the non-equilibrium steady state. Section 5 concludes.

## 2. Steady state barrier crossing

The Kramers barrier crossing problem considers the time evolution of a single particle distribution function $P(x,v,t)$, governed by the the (markovian) Fokker-Planck equation

$$\frac{\partial P(x,v,t)}{\partial t} = \frac{1}{m}\frac{\partial V(x)}{\partial x}\frac{\partial P}{\partial v} - v\frac{\partial P}{\partial x} + \gamma\frac{\partial}{\partial v}\left[vP(x,v,t) + \frac{k_B T}{m}\frac{\partial P}{\partial v}\right] \quad (1)$$

where $\gamma$ is the friction coefficient, $T$ is the temperature and $k_B$ is the Boltzmann constant. $V(x)$ is the barrier potential, which near the top may be represented by the inverted parabola,

$$V(x) \approx -\frac{1}{2}m\omega_B^2 x^2. \quad (2)$$

We consider a non-equilibrium steady state characterized by a thermal flux across the barrier, driven by a chemical potential bias between the left and right sides, quantified by the boundary conditions imposed on the distribution $P(x,v,t)$:

$$P(x,v,t) \xrightarrow{x\to-\infty} \lambda P_{eq}(x,v); \quad P_{ss}(x,v,t) \xrightarrow{x\to\infty} (1-\lambda)P_{eq}(x,v), \quad (3)$$

where $0 \leq \lambda \leq 1$ and

$$P_{eq}(x,v) \equiv P_B \exp\left(-\beta\left[(1/2)mv^2 + V(x)\right]\right); \quad \beta = (k_B T)^{-1} \quad (4)$$

where $P_B$ is the equilibrium probability density for $v=0$ at the barrier top. This normalization parameter will not affect our results. An explicit expression for the steady state distribution is given by

$$P_{ss}(x,v) = \lambda P_{SS}^{L\to R}(x,v) + (1-\lambda)P_{SS}^{R\to L}(x,v), \quad (5)$$

where $P^{L\to R}(x,v)$ and $P^{R\to L}(x,v)$ are steady state solutions of Eq. (1) that satisfy the boundary condition (3) with $\lambda = 0$ and $\lambda = 1$, respectively. These solutions were found by Kramers:[24]

$$P_{ss}(x,v)^{K \to K'} = 2P_{eq}(x,v) f^{K \to K'}(x,v); \qquad K, K' = L, R \qquad (6a)$$

$$f^{L \to R}(x,v) = \sqrt{\frac{\alpha m}{2\pi k_B T}} \int_{-\infty}^{v+\Gamma x} dz \exp\left(-\frac{\alpha m z^2}{2 k_B T}\right), \qquad (6b)$$

$$f^{R \to L}(x,v) = f^{L \to R}(-x,-v) = \sqrt{\frac{\alpha m}{2\pi k_B T}} \int_{v+\Gamma x}^{\infty} dz \exp\left(-\frac{\alpha m z^2}{2 k_B T}\right) \qquad (6c)$$

where

$$\Gamma \equiv -\frac{\gamma}{2} - \sqrt{\left(\frac{\gamma}{2}\right)^2 + \omega_B^2} \quad \text{or} \quad \frac{\Gamma}{\omega_B} = -\bar{\gamma} - \sqrt{\bar{\gamma}^2 + 1} \qquad (7)$$

$$\alpha \equiv -\frac{1}{2} + \sqrt{\frac{1}{4} + \left(\frac{\omega_B}{\gamma}\right)^2} = \frac{1}{2\bar{\gamma}}(\sqrt{\bar{\gamma}^2 + 1} - \bar{\gamma}); \quad \bar{\gamma} \equiv \gamma/(2\omega_B) \qquad (8)$$

Note that $f^{L \to R}(x,v) \to 1$ and $0$ when $x \to -\infty$ and $\infty$, respectively, and that $f^{L \to R}(x,v) + f^{R \to L}(x,v) = 1$. The latter idenstity implies that $P_{ss}(x,v;\lambda = 1/2) = P_{eq}(x,v)$.

## 3. Non-equilibrium barrier dynamics

An important case of the Kramers theory of barrier crossing is the high barrier limit where the reaction, i.e. barrier crossing, is slow relative to the rate of thermal relaxation in the well. In this limit, the barrier crossing rate is determined by the steady state flux associated with the distribution (5),[24] where, for example, $\lambda = 1$ if the reactant is represented by the population left of the barrier. It is in this barrier region where the system is out of thermal equilibrium and exchanges net heat with its environment.

The non-equilibrium character of the distribution $P_{ss}(x,v)$ near the barrier can be characterized by the local effective temperature, $T_{eff}(x)$, defined by the local kinetic energy,

$$T_{\text{eff}}(x) = \frac{m\langle v^2(x)\rangle}{k_B} \tag{9}$$

where

$$\langle v^2(x)\rangle = \frac{\int dv\, v^2 P_{ss}(x,v)}{\int dv\, P_{ss}(x,v)} \tag{10}$$

Figure 1 shows, for the case $\lambda = 1$, the ratio $T_{\text{eff}}(\bar{x})/T$ as a function of the dimensionless position $\bar{x} \equiv x(m\omega_B^2/k_B T)^{1/2}$ for different value of the dimensionless friction $\bar{\gamma} = \gamma/2\omega_B$. As expected, the $T_{\text{eff}}(x)$ is smaller than the ambient temperature at the climbing-up section (later referred to as the cold section) of the flux trajectory, $\bar{x} < 0$, and is larger than ambient for $\bar{x} > 0$. Deviation from equlibrium is larger for smaller $\gamma$ and vanishes in the source region $x \to -\infty$.

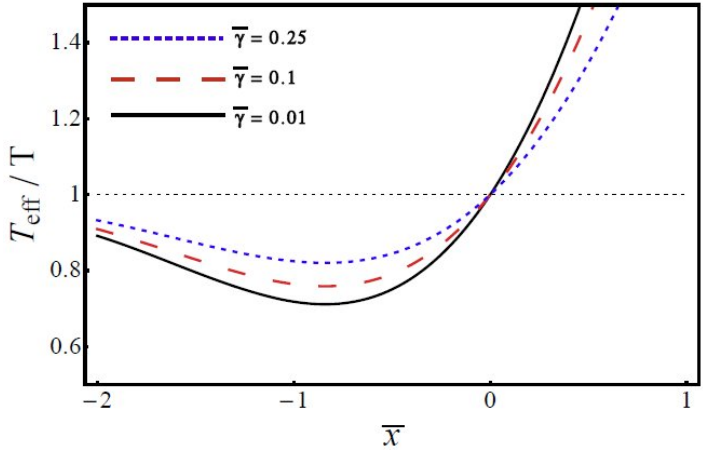

Figure 1: Effective temperature as a function of position in the steady state of the barrier crossing dynamics defined by Eqs. (1)-(5) with $\lambda = 1$. Results are displayed for several values of the dimensionless friction $\bar{\gamma} \equiv \gamma/2\omega_B$. $T_{\text{eff}}/T = 1$ is marked by a thin dotted line

The fact that in the cold section of the flux trajectory the system is colder than its thermal environment implies that in that region the system absorbs heat this environment. The local rate at which this cooling takes place, can be calculated from Eq. (1) with the steady state distribution $P_{ss}$. To this end consider the rate at which the average system

energy density (energy per unit length) $\rho_E(x)$ changes at position $x$

$$\left(\frac{d\rho_E(x,t)}{dt}\right) = \frac{d}{dt}\int_{-\infty}^{\infty} dv E(x,v) P(x,v,t) = \int_{-\infty}^{\infty} dv E(x,v)\left(\frac{dP(x,v,t)}{dt}\right) \quad (11)$$

where

$$E(x,v) = \frac{1}{2}mv^2 + V(x) \quad (12)$$

At steady state, $d\rho_E(x,t)/dt$ vanishes. However from Eqs. (1), (11) and (15), it can be written as a sum of non-zero deterministic and dissipative contributions that mutually cancel. In particular, the dissipative term, i.e. the contribution to $d\rho_E(x,t)/dt$ due to energy exchange with the thermal environment is given at steady state by

$$\left(\frac{d\rho_E(x,t)}{dt}\right)_{dissip}^{(ss)} = \gamma \int_{-\infty}^{\infty} dv E(x,v) \frac{\partial}{\partial v}\left[vP_{ss}(x,v) + \frac{k_B T}{m}\frac{\partial P_{ss}(x,v)}{\partial v}\right] \quad (13)$$

For our model, this rate can be evaluated analytically (see Appendix):

$$\left(\frac{d\rho_E(x,t)}{dt}\right)_{dissip}^{(ss)} = -4(2\lambda-1)\gamma|\Gamma|\left(\frac{\alpha}{1+\alpha}\right)^{3/2} P_B k_B Tx \quad (14)$$

It is useful to re-express this rate in terms the rate of energy change per particle at $x$

$$\left(\frac{d\varepsilon(x,t)}{dt}\right)_{dissip}^{(ss)} = \frac{1}{\rho_{ss}(x)}\left(\frac{d\rho_E(x,t)}{dt}\right)_{dissip}^{(ss)} \quad (15)$$

$$\rho_{ss}(x) = \int_{-\infty}^{\infty} dv P_{ss}(x,v) \quad (16)$$

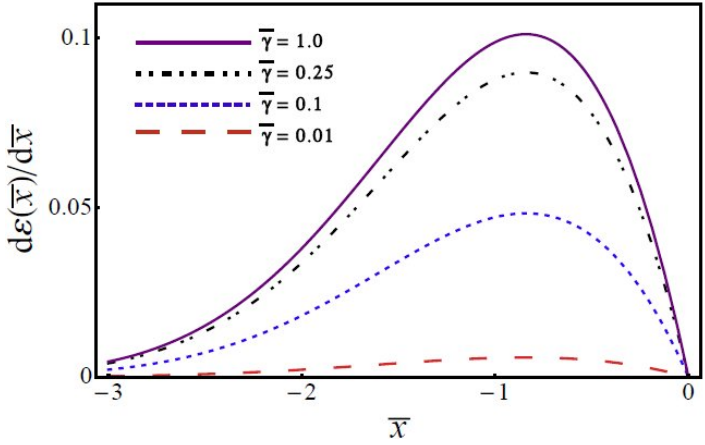

Figure 2: Energy exchange per particle, $(d\bar{\varepsilon}(x,t)/dt)_{dissip}$, displayed vs. $\bar{x}$ in the cold section of the flux trajectory, $\bar{x} < 0$, for $\lambda = 1$ and for different values of the friction $\bar{\gamma}$.

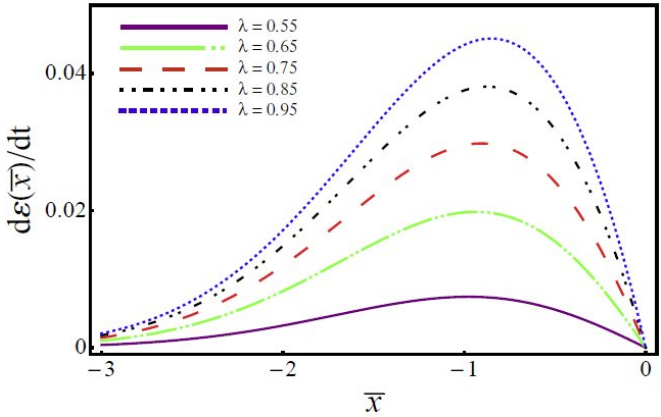

Figure 3: Same as Fig. 2. $(d\bar{\varepsilon}(x,t)/dt)_{dissip}$ is plotted against $\bar{x}$ for different values of the driving parameter $\lambda$. $\bar{\gamma} = \gamma/(2\omega_B) = 0.1$.

Figures 2 and 3 show this rate of energy exchange with the thermal environment per particle, $(d\bar{\varepsilon}(x,t)/dt)_{dissip}$, $\bar{\varepsilon} = \varepsilon/(k_B T)$, as a function of position in the cold region, $x < 0$, the friction, $\gamma$, and the driving parameter $\lambda$. $d\varepsilon(x,t)/dt > 0$ implies that the system absorbs heat from the environment in section of its crossing path. This remains true for $1/2 < \lambda \leq 1$, i.e. when the net flux across the barrier is from left to right.

The fact that the energy change rate per particle, $(d\bar{\varepsilon}(x,t)/dt)_{dissip}$ goes through a maximum as a function of *x* in the uphill section of the crossing path mimicss the minimum in the effective local temparture seen in Fig. 1. These extrema reflect the fact that, on one hand, the system approaches thermal equilibrium, i.e. vanishing net energy exchange with the environment, when $x \to -\infty$, and on the other the motion loses its uphill character as $x \to 0$.

In spite of this behavior of the energy exchange rate per particle, the rate of change in energy density (Eq. (14)) is linear in *x*, changing sign as expected at $x = 0$. This results from the exponential increase in the particle density as we go deeper into the wells, and constitutes an artifact of the bottomless parabolic barrier. This makes it necessary to introduce a cutoff energy in the model used in the next section to calculate the coefficient of performance of this setup, when used as a cooling machine.

## 4. Coefficient of performance

The analysis of Section 3 is based on Eq. (1) which is a phenomenologial stochastic equation describing the time evolution of a system coupled to a single heat bath. To view the system as a cooling engine one has to assume that it is possible to couple one heat bath locally to the system for $x < 0$ and another for $x < 0$, so that a driven barrier crossing process pumps heat from one bath to the other. The following analysis is based on this assumption. It should be emphasized that we did not derive Eq. (1) for such a model, and an attempt to do this will likely results in interface terms that are disregarded here. The following should be therefore considered as an heuristic consideration that serves to demonstrate the principle of heat pumping by a driven process with an uphill segment, rather than an exact model of such a machine.

The coefficient of performance (COP) of a heat pump is the ratio between the heat exchange with the reservoir of interest and the work input into the pump. It is sufficient to consider the range $1/2 < \lambda \leq 1$ where the net particle flux across the barrier is from left to right. In what follows we assume that each side of the system is in its own thermal equilibrium for $E < -E_B$, i.e.

$$x < x_L = -x_R = -2E_B / m\omega_B^2 \qquad (17)$$

(see Fig. 4). $E_B$ is taken to be large enough relative to $k_B T$, so that the results of the previous sections (rigorously obtained for equilibrium boundary conditions at $\pm\infty$) hold. The left and right thermal equilibria are characterized by the same temperature, $T$, but different chemical potentials. The difference

$$\mu_L - \mu_R \equiv \Delta\mu_{LR} = k_B T \ln\left(\frac{\lambda}{1-\lambda}\right) \qquad (18)$$

provides the driving force for the ensuing flux across the barrier. We further assume that useful heat absorption takes place throughout the region $x_L < x < 0$ so that the rate of heat absorption is (using Eqs. (14) and (17))

$$\dot{Q} = \int_{x_L}^{0} dx \left(\frac{\partial \rho_E(x)}{\partial t}\right)^{ss}_{dissip} = 4(2\lambda-1)\gamma|\Gamma|\left(\frac{\alpha}{1+\alpha}\right)^{3/2} P_B \frac{k_B T E_B}{m\omega_B^2} \qquad (19)$$

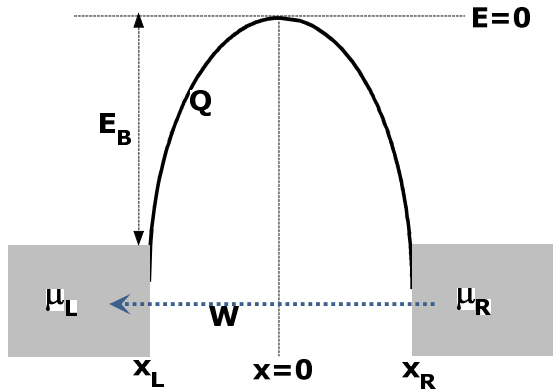

Figure 4. A schematic view of the Kramers heat pump.

The minimum work per unit time, $\dot{W}$, needed to maintain the steady state cooler operation is

$$\dot{W} = \Delta\mu_{LR}(2\lambda-1)J(\lambda=1) \qquad (20)$$

where $J(\lambda=1)$ is the steady state flux across the barrier associated with the distribution $P_{ss}^{L\to R}(x,v)$, Eq. (6). With our choice of normalization it is given by

$$J(\lambda=1) = \int_{-\infty}^{\infty} dv\, v\, P_{ss}^{L\to R}(x,v) = \frac{k_B T}{m}\left(\frac{\alpha}{\alpha+1}\right)^{1/2} P_B \qquad (21)$$

Eqs. (19)-(21) finally give the coefficient of performaance in the form

$$\eta = \frac{\dot{Q}}{\dot{W}} = 4\frac{\gamma|\Gamma|}{\omega_B^2}\frac{E_B}{k_B T}\frac{\alpha}{\alpha+1}\frac{1}{\ln\left(\frac{\lambda}{1-\lambda}\right)} \qquad (22)$$

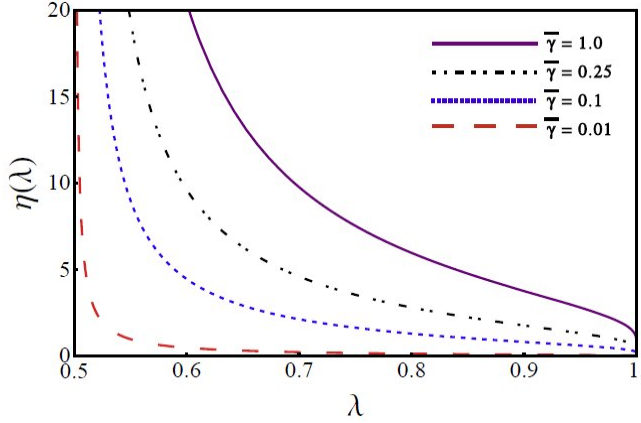

Figure 5: The COP, $\eta$, Eq. (22), shown as a function of the driving parameter $\lambda$ ($0.5 \leq \lambda \leq 1$), for different values of the friction $\bar{\gamma}$.

The following points are noteworthy: First, the COP vanishes for extereme driving, $\lambda = 0, 1$, becomes larger for a system closer to equilibrium. and diverges as $\lambda \to 0.5$. This behavior results from the $\lambda$ dependence of the chemical potential difference, Eq. (18). Secondly, $\eta$ increases with increasing friction $\bar{\gamma}$. This may appear surprising, since cooling is associated with the non-equilibrium distribution of the crossing particles near the barrier top, and it is for small $\gamma$ that this non-equilibrium distribution is most pronounced. Obviously, $\eta \to 0$ when $\gamma \to 0$ because heat exchange with the thermal environment vanishes in this limit. It is easy to show that $\eta$ becomes independent of $\gamma$ as $\gamma \to \infty$.

Finally, the linear dependence of the COP on the barrier height $E_B$ results from the particular structure of our parabolic barrier and the cutoff used, and should not be regarded as a generic property of this type of processes. On the other hand, it may be expected that $\eta$ will usually increase with the barrier height, since the latter determines the amount of energy needed for the uphill step that may be drawn from the thermal environment.

## 5. Summary and conclusion

Driven processes in which an intermediate uphill step is locally coupled to an external heat source can be used to cool this source. In this paper we have analyzed a simple example, a one dimensional classical barrier crossing process, and evaluated its properties as a cooling machine. Such analysis should be useful in exploring generic properties of this type of processes.

**Acklowledgment.** This paper is dedicated to Peter Hanggi, a collegue and a leader who has put his mark on our field for over three decades. This research was supported by the European Science Council (FP7 /ERC grant no. 226628), the German-Israel Foundation, and the Israel Science Foundation.

## Appendix: Evaluation of Eq. (13)

To evaluate the integral in (13) start with the case $\lambda = 1$ so that (cf. Eq. (6)) $P_{ss}(x,v) = P_{ss}(x,v)^{L \to R} = 2P_{eq}(x,v) f^{L \to R}(x,v)$. For this case

$$B(x,v) \equiv vP_{ss} + (k_B T/m)\partial P_{ss}/\partial v = \sqrt{\frac{2\alpha k_B T}{\pi m}} P_{eq}(x,v) e^{-(1/2)\alpha m (v+\Gamma x)^2/(k_B T)}$$

The needed integral can be evaluated by parts

$$\left(\frac{\partial \rho(x,t)}{\partial t}\right)^{(ss)}_{dissip} = \gamma \int_{-\infty}^{\infty} dv E(x,v) \frac{\partial}{\partial v} B(x,v) = -m\gamma \int_{-\infty}^{\infty} dv\, v\, B(x,v)$$

followed by a staightforward evaluation of the gaussian integral to yield the $\lambda = 1$ limit of Eq. (14). In the general case

$$\left(\frac{d\rho_E(x,t)}{dt}\right)^{(ss)}_{dissip} = \lambda \left(\frac{d\rho_E(x,t;\lambda=1)}{dt}\right)^{(ss)}_{dissip} + (1-\lambda)\left(\frac{d\rho_E(x,t;\lambda=0)}{dt}\right)^{(ss)}_{dissip}$$

and using

$$\left(\frac{d\rho_E(x,t;\lambda=1)}{dt}\right)^{(ss)}_{dissip} = \left(\frac{d\rho_E(-x,t;\lambda=0)}{dt}\right)^{(ss)}_{dissip}$$

Leads to Eq. (14).